\documentclass[useAMS,usenatbib]{mn2e}
\usepackage{graphicx}
\usepackage{txfonts}
\usepackage{diagbox}
\usepackage{threeparttable}
\usepackage{natbib}
\def\ms{{\rm M_{\odot}}}
\def\be{\begin{equation}}
\def\ee{\end{equation}}
\def\bea{\begin{eqnarray}}
\def\eea{\end{eqnarray}}

\def\ms{\ifmmode M_{\odot} \else $M_{\odot}$\fi}    

\def\m18{{\rm{\dot M_{18}}}}
\def\1808{\ifmmode {SAX J1808.4-3658 } \else SAX J1808.4-3658\fi}

\title[Minimum magnetic field of MSPs]
{The minimum magnetic field of millisecond pulsars by accretion: application to X-ray neutron star SAX J1808.4-3658 in LMXB }

\author[Y. Y. Pan et al.]{Pan Y. Y.$^{1, 2}$ \thanks{E-mail:
panyy@xtu.edu.cn, zhangcm@bao.ac.cn},
Zhang C. M.$^{2, 3, 4}$, Song L. M.$^{5}$, Wang N.$^{6}$, Li D.$^{2,  3}$, Yang Y. Y.$^{2,3,7}$\\
$^{1}$Department of Physics, Xiangtan University, Hunan, 411105, China\\
$^{2}$ National Astronomical
Observatories, Chinese Academy of Sciences, Beijing 100101, China\\
$^3$ CAS Key Laboratory of FAST,  Chinese Academy of Sciences, Beijing 100101, China\\
$^4$ School of Physical Sciences, University of Chinese Academy of Sciences, Beijing 101400, China\\
$^5$Key Laboratory of Particle Astrophysics, Institute of High Energy Physics, Chinese Academy of Sciences, Beijing 100049, China\\
$^6$ Xinjiang Astronomical Observatory, Chinese Academy of Sciences, Urumqi 830011, China\\
$^7$ Astronomy Department, Beijing Normal University, Beijing 100875, China}
\begin{document}
\date{Accepted Date; Received Date}

\pagerange{\pageref{firstpage}--\pageref{lastpage}} \pubyear{}

\maketitle

\label{firstpage}

\begin{abstract}
Based on the model of accretion induced the magnetic field decay of the neutron star (NS), the millisecond pulsars (MSPs) will own the minimum magnetic field when the NS magnetosphere radius shrinks to the stellar surface during the binary accretion phase. We find that this minimum magnetic field is related to the accretion rate $\dot{M}$ as $B_{\rm min}\sim2.0\times10^7{\,\rm G}\,(\dot{M}/\dot{M}_{\rm min})^{1/2}$, 
where $\dot{M}_{\rm min}=4.6\times10^{15}\rm\,g/s$ is the averaged minimum accretion rate required for the MSP formation and constrained by the long-term accretion time, which corresponds to the companion lifetime less than the Hubble time. The value of $B_{\rm min}$ is consistent with that of the observed radio MSPs and the accreting MSPs in low mass X-ray binaries, which can be found the case of the application on the minimum and present field strength of \1808. The prediction on the minimum magnetic field of MSPs would be the lowest field strength of NSs in universe, which could constrain the evolution mechanism of the magnetic field of accreting NSs.

\end{abstract}

\begin{keywords}
accretion: accretion disks --
binaries: close --
X-rays: stars--
stars: millisecond pulsars
\end{keywords}

\section{Introduction}
\begin{figure}
\includegraphics[width=8cm]{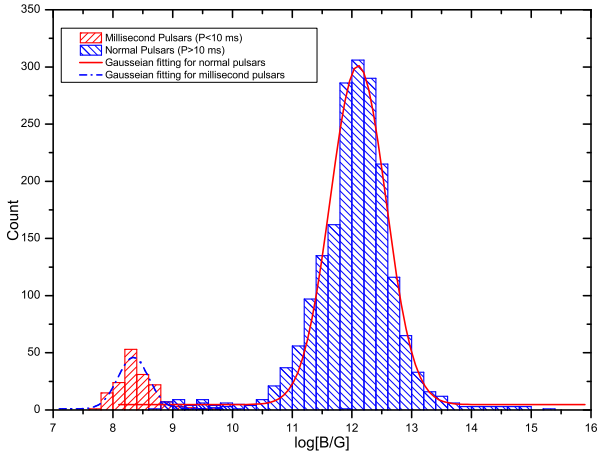}
\caption{Histogram of magnetic fields of 2636 pulsars (data from ATNF pulsar catalogue until April of 2018). The samples almost follow a bimodal distribution (Camilo et al. 1994): centered at $\sim10^{12.5}\rm\,G$  for normal pulsars labelled with the left-inclined bars and $10^{8.5}\rm\,G$ for MSPs labelled with  the right-inclined bars. }
\label{hist_b}
\end{figure}

Millisecond pulsars (MSPs) are with the spin period ($P$) less than $10\rm\,ms$ and the magnetic field ($B$) around {\bf $10^{8.5}\rm\,G$}. They are recycled pulsars (PSRs) through the accretion in the low mass X-ray binaries (LMXBs) (Stairs 2004; Lorimer 2008; van den Heuvel 2009, 2017; Manchester 2017). Since the first MSP (PSR B1937+21) discovered by Backer et al. in 1982, there have been over 300 MSPs recorded in ATNF pulsar catalogue until April of 2018 (Manchester et al. 2005). Their magnetic fields ($B$) are believed to decay from $\sim 10^{10.5-15.5}$ G (Ho 2013; Luo et al. 2015; Kaspi 2017) to $\sim10^{7.5-9.0}\rm\,G$ in LMXBs by accreting the mass of  $\sim0.1-0.2\ms$, which can be inferred from the observations (Bhattacharya \& van den Heuvel 1991; Phinney \& Kulkarni 1994; Bhattacharya \& Srinicasan 1995; van den Heuvel 2004; Ruderman 2010; Zhang et al. 2011, 2016; Tauris 2015; Manchester 2017). As a  comparison, the histogram of field strength of MSPs and normal PSRs are shown in Fig.\ref{hist_b}.

The model of a MSP formed during the accretion in LMXB was first proposed in 1980s (Alpar et al. 1982). Then, from the observational statistics, Taam and van den Heuvel (1986) found that the magnetic field of neutron star (NS) decayed inversely with the accretion mass, based on which, an empirical formula about the field strength evolution of the X-ray PSR with the accretion mass was presented by Shibazaki et al. (1989). Furthermore, by assuming the frozen magnetic field in the NS crust, the model of the accretion induced the magnetic field decay was proposed (Zhang \& Kojima 2006), which was applied to simulate the B-P evolution of accreting NSs, whose results were suitable for those of the observed PSRs (Wang et al. 2011; Pan et al. 2013, 2015).

The magnetic field of MSPs can be estimated by different methods. For the radio and non-accreting MSPs, their magnetic fields can be calculated  through the spin period and period derivative ($\dot{P}$): $B\simeq3.2\times10^{19}{\,\rm (G)}\,(P\dot{P})^{1/2}$ (Shapiro \& Teukolsky 1983; Bhattacharya \& van den Heuvel 1991).
For the anomalous X-ray PSRs, the magnetic field can be obtained through modeling the non-thermal X-ray spectra with cyclotron and magnetic Compton scattering processes in the magnetosphere (G$\rm\ddot{u}$ver, $\rm\ddot{O}$zel \& G$\rm\ddot{o}\breve{g}\ddot{u}s$ 2008). For the X-ray PSRs with high B (e.g. $B\geqslant10^{12} \rm\,G$), one can use the resonant electron cyclotron lines in the X-ray spectra (Caballero \& Wilms 2012).
For the accreting X-ray MSPs (AMXPs), the magnetic field can be derived through comparing the magnetosphere radius to the co-rotation radius of accreting NS (Burderi et al. 1996, 2002), by which the magnetic fields of \1808 (Wijnands \& van der Klis 1998) and XTE J1751-305 and XTE J0929-314 were estimated in the order of $\sim10^{8}$ G (Wijnands et al. 2005).
The magnetic field of a NS in LMXB (NS/LMXB) can also be constrained with its detected frequency of kilo-Hertz quasi-periodic oscillation (kHz QPO, van der Klis 2000; Zhang 2004).
In addition, if the frequency derivative of an accreting NS/LMXB is detected, the magnetic field can be estimated by the spin up formula of X-ray NS (Ghosh \& Lamb 1979), by which the magnetic fields of a dozen NS/LMXBs are obtained as $10^{8-9}\rm\,G$ with their frequency derivatives about $\sim10^{-14}\rm\,Hz\,s^{-1}$ (Burderi \& Di Salvo 2013).

In this paper, by the model of accretion induced the magnetic field decay of NS, we study the minimum field strength $B_{\rm min}$ of MSPs, which corresponds to the minimum accretion rate for the MSP formation in section 2. The consistence is investigated between our $B_{\rm min}$ of MSPs and that of the observed MSPs and AMXPs. We also figure out the magnetic field range of NS/LMXB whose companion is with the possible upper limit mass. Furthermore, the minimum magnetic field of AMXPs is discussed, including an application for the minimum and present field strength of the first discovered AMXP, \1808. The conclusion and discussion are given in section 3.

\begin{figure}
\includegraphics[width=8cm]{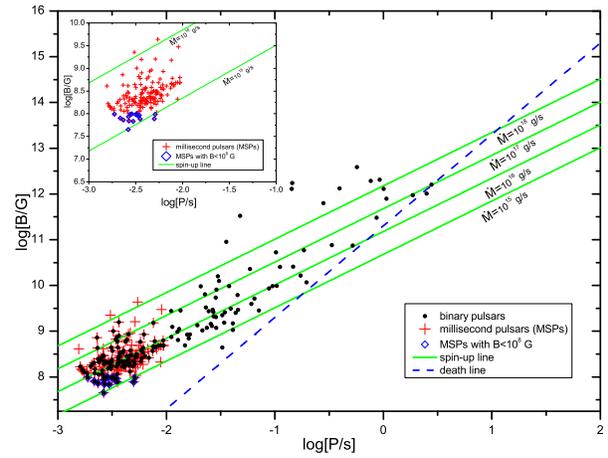}
\caption{Binary pulsars and millisecond pulsars (labelled with the dot and crossing symbols) distributed in the magnetic field versus spin period diagram. The ones with $B<10^8\rm\,G$ are labelled with the quadrangles. Four solid lines represent the spin up lines with the accretion rates from $10^{18}\rm\,g/s$ (upper) to $10^{15}\rm\,g/s$ (bottom), and  the dash line stands for the death line (Bhattacharya \& van den Heuvel 1991). The upper-left small B-P diagram is the enlarged close-up of MSPs. }
\label{msp}
\end{figure}

\section{Minimum magnetic field of accreting MSP}

\subsection{Accretion induced the magnetic field decay}

During the binary accretion phase, a NS can evolve to be a MSP by accreting  mass of $\sim 0.1-0.2\ms$ at least.
%
%
With such accreted mass, the bottom magnetic field ($B_{\rm f}$) of NS can be achieved when its magnetosphere radius shrinks from about a few thousand kilo-meters to the NS radius (van den Heuvel \& Bitzaraki 1995). %
The estimated work of $B_{\rm f}$ was obtained by the model of the accretion induced field strength decay of NS (Zhang \& Kojima 2006):
\begin{equation}
B_{\rm f}=1.32\times10^8{\rm\,(G)}\,(\frac{\dot{M}}{\dot{M}_{\rm Edd}})^{1/2}m^{1/4}R^{-5/4}_6\phi^{-7/4},
\label{bf}
\end{equation}
where $\dot{M}$ is the accretion rate in units of the Eddington accretion rate $\dot{M}_{\rm\,Edd}$, $m$ and $R_6$ are the NS mass $M$ and radius $R$  in units of solar mass and $10^6\rm\,cm$, respectively. The parameter $\phi$, always taken to be 0.5 (Li \& Wang 1999; Ghosh \& Lamb 1992), is a model dependent ratio between  the magnetosphere radius $R_{\rm M}$ and Alfv\'en radius $R_{\rm A}$ of NS (Shapiro \& Teukolsky 1983; Frank et al. 2002):
\begin{equation}
R_{\rm M}=\phi R_{\rm A},
\label{rm}
\end{equation}
\begin{equation}
R_{\rm A}=1.7\times10^8\,\,({\rm cm})\,\,(\frac{\dot{M}}{\dot{M}_{Ed}})^{-2/7}\mu_{30}^{4/7}m^{-1/7},
\label{ra}
\end{equation}
where $\mu_{30}=B_{12}R_6^3$ is the magnetic moment $\mu$ in units of $10^{30}\rm\,G\,cm^3$. $B_{12}$ is the magnetic field $B$ in units of $10^{12}\rm\,G$.

As can be seen from Eq. (\ref{bf}), the bottom magnetic field $B_{\rm f}$, which is also the minimum magnetic field of NS, is mainly depended on the accretion rate, and little affected by the NS mass and radius. During the NS evolution, if the minimum accretion rate can be deduced, its minimum magnetic field will be acquired. Thus, in the following subsection we will discuss the minimum accretion rate for the MSP formation and the corresponding magnetic field.

\subsection{The minimum accretion rate and the corresponding magnetic field of NS}

In the binary system, the accretion rate related to the accretion mass $\Delta M$ and accretion time $t_{\rm ac}$ is
\begin{equation}
\dot{M} = \frac{\Delta M}{t_{\rm ac}},
\label{mdot_ac}
\end{equation}
where $t_{\rm ac}$ is associated with the
age of the companion in the main sequence time $T_{\rm MS}$ (Shapiro \& Teukolsky 1989):
\begin{equation}
t_{\rm ac}=\zeta\, T_{\rm MS},
\label{t_ac}
\end{equation}
\begin{equation}
T_{\rm MS}\simeq1.3\times10^{10}\,\,{\rm(yr)}\,\,m_{\rm c}^{-2.5},
\label{t_hubble}
\end{equation}
the parameter $\zeta$ is usually taken as 10\% (Shapiro \& Teukolsky 1989), and $m_{\rm c}=M_{\rm c}/\ms$ is the companion mass $M_{\rm c}$ in units of solar mass. If $T_{\rm MS}$ is closed to the Hubble age or the age of universe: $t_{\rm H}\simeq1.38\times10^{10}\rm\,yr$ (Planck et al. 2016), there will exist a companion star in the main sequence with the minimum mass of about $1.0\ms$ according to Eq. (\ref{t_hubble}). Such result is consistent with the minimum companion mass of $0.8\ms$ in LMXB estimated by van den Heuvel \& Bitzaraki (1995). When the least accretion mass of $\sim 0.1\ms$ of producing a MSP is acquired by the NS in LMXB, there will be the longest accretion time, which can result in a minimum accretion rate $\dot M_{\rm min}$ with Eqs. (\ref{t_ac}) and (\ref{t_hubble}). After the arrangement, the accretion rate $\dot M$ can be written with the minimum accretion rate $\dot M_{\rm min}$:
\begin{equation}
\dot{M} = \dot{M}_{\rm min}\,m_{\rm c}^{2.5}\,(\frac{\Delta{M}}{0.1\ms})\,(\frac{\zeta}{0.1})^{-1},
\label{mdot_ac1}
\end{equation}

\be
\dot{M}_{\rm min}=4.6\times10^{15}\rm\,g/s.
\ee
If $\dot M<\dot{M}_{\rm min}$, the NS could not evolve to a MSP in LMXB. Thus, we call $\dot{M}_{\rm min}$ to be the critical accretion rate for the MSP formation, which can be found  the consistency  with the B-P distribution of MSPs, as shown  in Fig. \ref{msp}:
MSPs are almost above the spin-up line with $\dot M=10^{15}\,\rm g/s$ and  approximately gathered around the spin-up line with $\dot M=10^{16}\,\rm g/s$. This scenario illustrates the accretion rate requested for the MSP formation is closed to the theory result of $\dot M_{\rm min}=4.6\times10^{15}\rm\,g/s$, since the current B-P distribution of MSPs in Fig. \ref{msp} should be similar to that of their birth positions (Camilo et al. 1994).

Thus the minimum magnetic field of MSP can be expressed by Eq.
(\ref{bf}) with $\dot M_{\rm min}$ as:
\begin{equation}
B_{\rm min}\simeq2.0\times10^7{\,\rm (G)}\,(\frac{\dot{M}}{\dot{M}_{\rm min}})^{1/2} (\frac{m}{1.4})^{1/4}(\frac{R_6}{1.5})^{-5/4}(\frac{\phi}{0.5})^{-7/4},
\label{bmin}
\end{equation}
which shows a simply relation to the accretion rate $B_{\rm min}\sim2.0\times10^7{\,\rm (G)}\,(\dot{M}/\dot{M}_{\rm min})^{1/2}$.
When taken the critical accretion rate $\dot{M}_{\rm min}$ into consideration, the minimum magnetic field of all MSPs for conditions $m=1.4$ and $R_6=1.5$ is $B_{\rm min}\simeq2.0\times10^7\rm\,G$. It might also be the minimum magnetic field of all NSs in the universe. We list the minimum field strengths of MSPs in Table \ref{bf_difr} that are influenced by the selected parameters: the accretion mass, NS radius and accretion rate.

As a comparison, the 16 MSPs with the magnetic fields lower than $10^8\rm\,G$ are listed in Table 2, where the parameters of $P$, $\dot P$ and $B$ are taken from ATNF pulsar catalogue (Manchester et al. 2005). From Table 2, we find that even the lowest  observed B of MSPs, e. g., $B=4.5\times10^7\rm\,G$ of PSR J1938+2021, is still higher than our calculated value of $B_{\rm min}$. 
It might be attributed to the fact that  the magnetosphere of PSR J1938+2021's has not been compressed onto the the stellar surface in  the accretion phase, otherwise  its current spin period would be about one millisecond (van der Klis 2000; Zhang 2004). 
What is more, the field strength of the observed radio MSPs in ATNF data catalogue are deduced by the magnetic dipole model, which would also lead to the difference from our calculated $B_{\rm min}$. For the AMXPs, the minimum field strengths of the 14 AMXPs calculated by Mukherjee et al. (2015) are also lower than our $B_{\rm min}$. The explanation might be that $B_{\rm min}$ of MSPs in this paper is based on the mean value of the minimum accretion rate during the accretion, and the work by  Mukherjee et al. (2015) was made  with the condition of the lowest X-ray luminosity of those 14 AMXPs.

\subsection{The field strength range of MSP corresponding to the upper limit mass of $M_{\rm c}$ in main sequence in LMXB}

During the accretion, the maximum accretion mass is expected to be captured by the NS/LMXB. However, a sizeable part of the companion mass loses from the system. Thus the accreted mass of NS will be a fraction to the companion mass with a ratio coefficient $f$:
\be
\Delta m=f\,m_{\rm c},
\ee
and the accretion rate will be:
\begin{equation}
\dot{M}=\frac{M_{\rm c}}{t_{\rm ac}}\simeq4.8\times10^{16}\,\,({\rm g/s})\,\,(\frac{\zeta}{0.1})^{-1}\,f\,m^{3.5}_{\rm c}.
\label{m_f}
\end{equation}
where $f$ may be in the order of 0.5 as argued by van den Heuvel and Bitzaraki (1995). Since the mass statistics of NSs and MSPs illustrates that the accreted mass is about $0.1-0.2 \ms$ for the MSP formation at least (Zhang et al. 2011; Kiziltan et al. 2013; Ozel \& Freire 2016; Antoniadis et al. 2016), we prefer $f$ to be about $0.1-0.2$.

A NS/LMXB with $m=1.4$ and $R_6=1$ for $M_{\rm c}>1.0\ms$ possesses the accretion rate larger than $\dot M_{\rm min}$. Meanwhile it cannot accrete material over Eddington-limiting, e.g., $\dot M\leq\dot{M}_{\rm Ed}=2\times 10^{18}\rm\,g/s$. With the limit of the accretion rate $\dot{M}=\dot{M}_{\rm Ed}$, one can derive an upper limit companion mass in the main sequence according to Eq. (\ref{m_f}):
\be
M_{\rm c}^{\rm max}\simeq5.6\ms\cdot(\frac{f}{0.1})^{-2/7}\;.
\ee
With $0.1<f<0.2$, $M_{\rm c}^{\rm max}$ will be about $4.6\ms$ to $5.6\ms$.
Hence a possible mass distribution of the companion in main sequence in LMXB is $1.0\ms\leq M_{\rm c}\leq M_{\rm c}^{\rm max}$, corresponding to the accretion rate of $\dot M_{\rm min}\leq\dot{M}\leq\dot{M}_{\rm Edd}$. In such a LXMB, the NS will evolve to be a MSP with the magnetic field of $2.0\times10^7\rm\,G\leq B\leq2.8\times10^8\rm\,G$. While for $M_{\rm c}>M_{\rm c}^{\rm max}$, the accretion time might be too short for the binary system to produce a MSP.

\begin{table}
\centering
\caption{The minimum  magnetic fields of MSPs with the different radii and accretion rates}
\begin{tabular}{|c|c|c|c|}
  \hline
  \diagbox{$\dot{M}\rm\,(10^{15}\rm\,g/s)$}{$B_f\,(10^7\rm\,G)$}{$R\rm\,(km)$}&15&10\\
  \hline
  4.6 ($\Delta M=0.1\ms$)&2.0&3.3\\
  9.2 ($\Delta M=0.2\ms$)&2.8&4.6\\
  \hline
\end{tabular}
\label{bf_difr}
\end{table}

\begin{table}
\centering
\label{ble8}
\caption{The 16 radio MSPs with the low magnetic fields of $\sim <10^8\rm\,G$ {\, \, \,  \, } (Data from ATNF pulsar catalogue) }
\begin{tabular}{@{}clccc@{}}
\hline
     No.& Name& $P$\,(${\rm ms}$) &$\dot P\,(10^{-21}\rm\, {s\,s^{-1}}$) &$B\,(10^7\rm\, G$)\\

  \hline

1&	J1938+2012&	2.63&       0.75& 4.50\\
2&	J2229+2643&	2.98&       1.52& 6.81\\	
3&	J1327-0755&	2.68&       1.77& 6.97\\	
4&	J1017-7156&	2.34&        2.22& 7.29\\	
5&	J1216-6410&	3.54&	1.62& 7.65\\	
6&	J0514-4002A&	4.99&	1.17& 7.73\\	
7&	J1544+4937&	2.16&	2.93& 8.05\\	
8& 	J1745+1017&	2.65&       2.73& 8.61\\	
9&	J1836-2354A&	3.35&	2.32& 8.92\\	
10&	J2317+1439&	3.45&	2.43& 9.26\\	
11&	J1640+2224&	3.16&	2.82& 9.55\\
12&	J1101-6424&	5.11&	1.80& 9.70\\	
13&	J1906+0055&	2.79&	3.32& 9.74\\	
14&	J0034-0534&	1.88&	4.97& 9.77\\	
15&	J1910-5959A&	3.27&	2.95& 9.93\\	
16&	J0636+5129&	2.87&       3.38& 9.96\\	

\hline
\end{tabular}
\end{table}

\subsection{On the minimum magnetic fields of AMXPs}

AMXPs exhibit the X-ray outburst during the accreting phase with the luminosity floating a few magnitudes higher than that in the quiescence phase (Wijnands \& van der Klis 1998; van der Klis 2000, 2006; Mukherjee et al. 2015). The variation of the luminosity leads to the changing of the accretion rate, thus the magnetic field of AMXPs can be derived by equating the magnetosphere radius with the co-rotation radius $R_{\rm co}$ of NS (Burderi et al. 1996; Zhang \& Kojima 2006):
\begin{equation}
B=B_{\rm f}\,(\frac{R_{\rm co}}{R})^{7/4},
\label{b_co}
\end{equation}
\begin{equation}
R_{\rm co}\simeq1.5\times10^{6}{\rm\,(cm)}\,m^{1/3}P^{2/3}_{-3},
\label{r_co}
\end{equation}
where $P_{-3}$ is the spin period $P$ of AMXPs in units of milliseconds. When $R_{\rm co}$ equals the NS radius, the AMXP will own the minimum magnetic field. For 19 AMXPs with the spin periods between $\sim1.7-6.1\rm\, ms$ (Patruno et al. 2017), their $R_{\rm co}$ are in the range of $\sim24-56\rm\,km$, which are larger than the radii of NSs of $\sim10-15\rm\,km$. Thus, it can be concluded that the magnetic fields of 19 AMXPs have not achieved their bottom values.

Another method for judging the minimum magnetic field of AMXPs is comparing their spin frequencies to the kilohertz quasi-periodic oscillation (kHz QPO) frequencies. The upper limit kHz QPO frequency of a AMSP is believed to be the Keplerian frequency with the orbital accreting matter (van der Klis 2000; Zhang 2004):
\begin{equation}
\nu_{\rm k}=(\frac{{\rm G}M}{4\pi^2R^3})^{1/2}\simeq1839{\rm\,(Hz)}\,(\frac{m}{R_6^3})^{1/2},
\label{fre}
\end{equation}
where $G$ is the gravitational constant. For the NS with $m=1.4$ and $R_6=1.5$, the orbital frequency $\nu_{\rm k}$ is $1184\rm\,Hz$. Some AMXPs are with their upper kHz QPO frequency over $1000\rm\,Hz$ (van der Klis 2000, 2006; Wang et al. 2017), which can be fairly compared with $\nu_{\rm k}$. However, the spin frequency of 19 AMXPs is from $164\rm\,Hz$ to $599\rm\,Hz$ (Patruno et al. 2017), which shows a deep gap to their upper kHz QPO frequency or $\nu_{\rm k}$, hence we conclude that the minimum magnetic fields of 19 AMXPs have not yet achieved. If the co-rotation radius approaches the stellar radius, AMXPs will own the minimum magnetic field through the accretion.

\subsection{The minimum magnetic field of SAX J1808.4-3658}

\1808 is the first AMXP discovered by Wijnands \& van der Klis (1998). Since its discovery, seven outbursts have been recorded until 2015 (Sanna et al. 2017). With the valuable observed information, we try to find the mean values of mass, radius and accretion rate to estimate the minimum magnetic field and present field value of \1808.

There are many mass researches on \1808. Leahy et al. (2008) constrained the source mass to be $1.3\ms$, that was almost consistent with the result $1.35\ms$ by Chakrabarty \& Morgan (1998). Elebert et al. (2009) derived the mass range of this source to be $0.6-1.8\ms$. While the latest research on source mass was about $0.97\ms$ (Wang et al. 2013), that is similar to the lower mass limit $\sim1.0\ms$ with the data of 1998, 2002 and 2005 outbursts proposed by Morsink \& Leahy 2011.
Based on these mass researches, a proximately mean value of mass is taken to be $1.3\ms$ for the magnetic field calculation of SAX J1808.4-3658.

Burderi and King (1998) calculated the NS radius of \1808 to be  $15\rm\,km$ according to the radius restriction $R \sim 13.8\,m^{1/3}\rm\,km$ with $m=1.3\ms$. While Papitto et al. (2009) proposed the radius to be $18\rm\,km$ with $m=1.4\ms$ by fitting with the disk-line profile, whose emission was identified from the neutral (or mildly ionized) iron. According to such radius results, the mean value of NS radius for \1808 is selected to be $16.5\rm\,km$.

Patrumo et al. (2017a) deduced the luminosity of \1808 was about $10^{36}\rm\,erg/s$, which was confirmed through the bolometric flux of the first five times outburst (Galloway \& Cumming 2006). Other researches on the source luminosity were $(4.7-6.4)\times10^{35}\rm\,erg/s$ during the short-live peak (Hartman et al. 2008) or $6.6\times10^{36}\rm\,erg/s$ with the source distance $3.5\rm\,kpc$ during the outburst stage (Hartman et al. 2008; Papitto et al. 2009). These results give the source a mean value of the luminosity to be $10^{36}\rm\,erg/s$, which corresponds to the accretion rate to be $10^{16}\rm\,g/s$.

Thus, with the mean values of $m=1.3$, $R_6=1.65$ and $\dot{M}=10^{16}\rm\,erg/s$ from the above discussion for \1808, we derive its minimum field strength $B_{\rm min}\simeq2.5\times10^7\rm\,G$ and present field strength  $B\simeq7.1\times10^7\rm\,G$, according to Eqs. (\ref{bmin}) and (\ref{b_co}). Its minimum field stength approaches to our result $2.0\times10^7\rm\,G$, and the present magnetic field approaches to $\sim10^8\rm\,G$ that was evaluated by Wijnands \& van der Klis (1998) and Chen (2017).
SAX J1808.4-3658 is still in the spinning-up phase (Di salvo \& Burderi 2003), and it will continue the field decay process until the magnetosphere collides its surface.

\section{Conclusion and discussion}

The minimum magnetic field of MSP appears when the NS magnetosphere is compressed onto the stellar surface, according to the model of  accretion induced the field strength decay. We find that the minimum field strength is proportionally related to the accretion rate, shown as $B_{\rm min}\simeq2.0\times10^7{\rm\, G}\,(\dot M/\dot M_{\rm min})$ for a MSP with $m=1.4$, $R_6=1.5$ and $\dot M_{\rm min}\simeq4.6\times10^{15} \rm\,g/s$.
$\dot M_{\rm min}$ is the critical minimum accretion rate for the MSP formation, which is constrained by two conditions: the accretion mass $0.1\ms$ required for the MSP formation at least, and the longest accretion time about 10\% of the main-sequence time of the companion star that closes to the Hubble age or the age of universe.

With the accretion rate $\dot M_{\rm min}$, the minimum magnetic field $2.0\times10^7\rm\,g/s$ of a MSP is obtained, that is closed to the minimum field strength of the observed MSPs, as shown in Fig. \ref{hist_b} and Table 2. Thus, we propose that the calculated minimum magnetic field of MSP is also the minimum field strength of NS, which would be a limitation of the magnetic field decay of NS in the universe. When the accretion rate is from $\dot M_{\rm min}$ to $\dot M_{\rm Edd}$ in LMXBs, the magnetic field of MSPs will be  $2.0\times10^7\rm\,G\leq B\leq2.8\times10^8\rm\,G$, which gives a requirement for the companions mass range in main sequence to be about $1.0\ms$ to $4.6$ or $5.6\ms$.

Some approximations exist during the estimation of $B_{\rm min}$, such as the accretion time and ratio between the accretion disk radius and Alfv\'en radius of the NS, which may cause $B_{\rm min}$ a little change:
(a) the accretion time is a fraction of the companion lifetime (10\%). The relaxation of this condition will modify $B_{\rm min}$;
(b) the ratio between the magnetic sphere and Alfv\'en radius is 0.5 as a usual choice (Ghosh \& Lamb 1979; Wang 1996; Ho et al. 2017). A higher one, e.g. $0.8\sim1$ (Li \& Wang 1999), could also arise $B_{\rm min}$.


When comparing $B_{\rm min}$ to the estimated field strengths of AMXPs, all AMXPs are with the higher magnetic field ($\sim 10^8 {\rm\, G}$) than the obtained minimum field value, e.g., the field strength calculation of the first AMXP, \1808. Its mean values of mass, radius and accretion rate of the source are selected based on the researches on its seven bursts in the past 20 years. With such conditions, the present magnetic field of the source is calculated to be $7.1\times10^7\rm\,G$, and the minimum value is $2.5\times10^7\rm\,G$. The results illustrate that \1808 will continue the evolution till the magnetic field decays to the minimum value.

\section*{Acknowledgement}
Thanks to Kulkarni S. R. for the discussions about this work.
This work is supported by the National Natural Science Foundation of China (Grant Nos. 11703021, 11173034, 11673023, 11690024, U1731238); the Strategic Priority Research Program on Space Science, the Chinese Academy of Sciences (Grant No. XDA 04010300); the National Program on Key Research and Development Project (Grant No. 2016YFA0400801); the Strategic Priority Research Program of the Chinese Academy of Sciences (Grant No. XDB23000000); Hunan Provincial NSF 2017JJ3310, 2018JJ3495; Guizhou Provincial Key Laboratory of Radio Astronomy and Data Processing (KF201817).

\end{document}